\documentclass[preprint,showpacs,preprintnumbers,amsmath,amssymb]{revtex4}

\usepackage{graphicx}
\usepackage{dcolumn}
\usepackage{bm}

\begin{document}


\title{Supersymmetry and Vector-like Extra Generation}

\author{Chun Liu}
\affiliation{Key Laboratory of Frontiers in Theoretical Physics and 
Kavli Institute for Theoretical Physics China, \\
Institute of Theoretical Physics,
Chinese Academy of Sciences,\\
P.O. Box 2735, Beijing 100190, China}
 \email{liuc@mail.itp.ac.cn}

\date{\today}

\begin{abstract}
Within the framework of supersymmetry, the particle content is extended 
in a way that each Higgs doublet is in a full generation.  Namely in 
addition to ordinary three generations, there is an extra vector-like 
generation, and it is the extra slepton SU(2)$_L$ doublets that are 
taken to be the two Higgs doublets.  R-parity violating interactions 
contain ordinary Yukawa interactions.  Breaking of supersymmetry and 
gauge symmetry are analyzed.  Fermion and boson spectra are calculated.  
Phenomenological constraints and relevant new physics at Large Hadron 
Collider are discussed. 
\end{abstract}



\maketitle

\section{Introduction}

The main stream spirit of the physics beyond the Standard Model (SM) is 
like the following.  SM gauge interactions unify at a high scale 
($\sim 10^{16}$ GeV) \cite{gut}.  Supersymmetry (SUSY) \cite{susy} is 
then a must to stabilize the SM Higgs mass.  The minimal SUSY extension 
of SM (MSSM) which necessarily involves two Higgs doublets, reinforces 
the idea of the grand unification theory (GUT) because of LEP data 
\cite{lep}.  Further assuming R-parity conservation, the dark matter 
(DM) is provided \cite{dm}.  

However, things can be different.  Before verification of SUSY GUT, 
other ideas should be pursued whenever they have reasonable points.  In 
this Large Hadron Collider (LHC) era, discoverable ideas besides MSSM 
are of particular interests.  

In this paper, we still use weak scale SUSY to extend the SM.  Our 
consideration is as follows.  The SUSY SM needs two Higgs doublets in 
order to provide masses to both up- and down-type quarks.  Fermions 
in each doublet cause anomaly.  In MSSM, anomalies of the two 
doublets would cancel each other.  The two Higgs 
doublets would be vector-like matter in MSSM.  They are very different 
from the three generation chiral matter in which the anomaly of each 
generation automatically vanishes.  To treat the Higgs and the matter 
equal footing, we introduce additional fields associated with each 
Higgs doublet, making them one generation leptons and quarks alike, the 
anomaly due to 
each SU(2) Higgsino cancels those of the new introduced.  In such an 
extension, for example, the down-type Higgs is in the same position as 
a lepton doublet in a full matter generation.  Immediately we find 
that except for the lepton and baryon numbers, the down-type Higgs and 
its associated fields can be identified as another full matter 
generation, that is the fourth generation.  

The fourth chiral generation may have been generally disfavored because 
of the $Z^0$ decays which shows three generation light neutrinos only.  
There were many discussions about the fourth generation 
\cite{vector,4th,gk,stu}.  Ref. \cite{gk} introduced the fourth 
generation and identified its slepton as the Higgs.  The partial role 
of sleptons in electro-weak symmetry breaking (EWSB) was discussed 
previously \cite{ls}.  Different from Ref. \cite{gk}, we take the weak 
scale being a low energy one, therefore we still have two Higgs 
doublets.  As we will see, whence introducing SUSY, the fourth generation 
neutrino is automatically heavy because of the existence of the 
generation associated with the up-type Higgs.  

The extra generations in this model are vector-like.  There are quite a 
few studies of vector-like fermionic generations \cite{vector,pati,nath}.  
Vector-like fermions may have interesting physical implications 
\cite{vf,decoupling}.  Within 
SUSY, Ref. \cite{nath} studied one mirror generation case.  We introduce 
a vector-like generation pair which contains the two Higgs doublets 
required for EWSB. 

This paper is organized as follows.  In Sect. II, the model is constructed.
SUSY breaking, EWSB and particle spectra are presented.  In Sect. III, 
phenomenological constraints are discussed.  LHC phenomenology is given in 
Sect. IV.  We summarize the model and further discuss its other aspects 
in the final section.

\section{Model}

In this paper, we consider a SUSY SM with a vector-like generation.  The 
particle contents are given bellow.  In addition to the fourth generation 
superfields, $L_4$, $E_4^c$, $Q_4$, $U_4^c$, $D_4^c$, the up-type Higgs 
$H_u$ and its associated matter ($E_H^c$, $Q_H$, $U_H^c$, $D_H^c$), which 
compose an anomaly-free 
chiral generation, are introduced.  Their quantum numbers under 
SU(2)$_L\times$U(1)$_Y\times$SU(3)$_c$ and the global baryon number are 
\begin{equation}
\begin{array}{c}
L_m(2,-1,1,0)\,, E_m^c(1, 2,1,0)\,, Q_m(2, \frac{1}{3},     3,\frac{1}{3})\,, 
U_m^c(1,-\frac{4}{3},\bar{3},-\frac{1}{3})\,, 
D_m^c(1, \frac{2}{3},\bar{3},-\frac{1}{3})\,, \\[3mm]
H_u(2, 1,1,0)\,, E_H^c(1,-2,1,0)\,, Q_H(2,-\frac{1}{3},\bar{3},-\frac{1}{3})\,, 
U_H^c(1, \frac{4}{3},     3,  \frac{1}{3})\,, 
D_H^c(1,-\frac{2}{3},     3,  \frac{1}{3})\,, 
\end{array}
\nonumber
\end{equation} 
where $m=1-4$.  In fact, the up-type Higgs family is in the anti-particle 
representation compared to particles in the other four ordinary generations.  
It will be massive after combining with one of the ordinary families.  

The superpotential is written as follows.  Instead of the R-parity, 
baryon number conservation is assumed, 
\begin{equation}
\label{1}
\begin{array}{lll}
{\mathcal W}&=& \mu_m L_m H_u + \mu_m^e E_m^c E_H^c + \mu_m^Q Q_m Q_H  
                + \mu_m^U U_m^c U_H^c + \mu_m^D D_m^c D_H^c 
                + \lambda_{lmn} L_l L_m E_n^c  \\[3mm]
            & & \lambda'_{lmn} Q_l L_m D_n^c 
                + y_{mn} Q_m H_u U_n^c + y'_m Q_H L_m U_H^c 
                + \tilde{y}_m E_H^c D_m^c U_H^c 
                + \tilde{y}_{mn} E_m^c D_H^c U_n^c \,,
\end{array} 
\end{equation} 
where $\mu_m$'s are mass parameters, $\lambda^{(\prime)}$, $y^{(\prime)}$ and 
$\tilde{y}^{(\prime)}$'s coefficients.  Note, $l, m, n= 1-4$.  By redefining 
the down-type Higgs and the other fourth generation fields, 
\begin{equation}
\label{2}
H_d\equiv \frac{\mu_m}{\mu}L_m\,,~~E_4^c\equiv\frac{\mu_m^e}{\mu^e}E_m^c\,,~~ 
Q_4\equiv\frac{\mu_m^Q}{\mu^Q}Q_m\,,~~U_4^c\equiv\frac{\mu_m^U}{\mu^U}U_m^c\,,
~~D_4^c\equiv\frac{\mu_m^D}{\mu^D}D_m^c\,,~~
\end{equation}
where 
\begin{equation}
\label{3}
\mu\equiv \sqrt{\sum_{m=1}^{4}|\mu_m|^2}\,,~~
\mu^e\equiv \sqrt{\sum_{m=1}^{4}|\mu_m^e|^2}\,,~~
\mu^Q\equiv \sqrt{\sum_{m=1}^{4}|\mu_m^Q|^2}\,,~~
\mu^U\equiv \sqrt{\sum_{m=1}^{4}|\mu_m^U|^2}\,,~~
\mu^D\equiv \sqrt{\sum_{m=1}^{4}|\mu_m^D|^2}\,,
\end{equation} 
the superpotential is  
\begin{equation}
\label{4}
\begin{array}{lll}
{\mathcal W}&=& \mu H_d H_u + \mu^e E_4^c E_H^c + \mu^Q Q_4 Q_H  
                + \mu^U U_4^c U_H^c + \mu^D D_4^c D_H^c 
                + y_{ij}^l L_i H_d E_j^c + y_{ij}^d Q_i H_d D_j^c\\
            & & + y_{ij}^u Q_i H_u U_j^c + \lambda_{ijk} L_i L_j E_k^c 
                + \lambda'_{ijk} Q_i L_j D_k^c 
                + \lambda_{ij}^E L_i L_j E_4^c + y_i^E L_i H_d E_4^c 
                + \lambda_{ij}^Q Q_4 L_i D_j^c \\
            & & + y_i^{Q\prime} Q_4 H_d D_i^c +\lambda_{ij}^DQ_iL_jD_4^c 
                + y_i^D Q_i H_d D_4^c + \lambda_i^{QD} Q_4 L_i D_4^c 
                + y^{QD} Q_4 H_d D_4^c + y_i^U Q_i H_u U_4^c \\
            & & + y_i^Q Q_4 H_u U_i^c + y^{QU} Q_4 H_u U_4^c 
                + \lambda'_i Q_H L_i U_H^c + y Q_H H_d U_H^c 
                + \tilde{\lambda}_i E_H^c D_i^c U_H^c
                + \tilde{\lambda} E_H^c D_4^c U_H^c\\
            & & + \tilde{\lambda}_{ij} E_i^c D_H^c U_j^c 
                + \tilde{\lambda}_i^U E_i^c D_H^c U_4^c 
                + \tilde{\lambda}_i^E E_4^c D_H^c U_i^c 
                + \tilde{\lambda}^{EU} E_4^c D_H^c U_4^c \,,
\end{array} 
\end{equation}
where field decomposition have been generally written as follows with 
$i$ being $1-3$, 
\begin{equation}
\label{5}
\begin{array}{c}
L_m   = c_{mi} L_i     + c_{m4} H_d     \,, ~~ 
E_m^c = c_{mi}^e E_i^c + c_{m4}^e E_4^c \,, ~~
Q_m   = c_{mi}^Q Q_i   + c_{m4}^Q Q_4   \,, \\
U_m^c = c_{mi}^U U_i^c + c_{m4}^U U_4^c \,, ~~
D_m^c = c_{mi}^D D_i^c + c_{m4}^D D_4^c \,, 
\end{array}
\end{equation} 
and the coefficients are 
\begin{equation}
\label{6}
\begin{array}{c}
y_{ij}^l = 2\lambda_{lmn} c_{li}c_{m4}c_{nj}^e \,,~~ 
y_{ij}^d = \lambda'_{lmn} c_{li}^Qc_{m4}c_{nj}^D \,,~~
y_{ij}^u = y_{mn} c_{mi}^Qc_{nj}^U \,,~~ 
\lambda_{ijk}  = \lambda_{lmn} c_{li}c_{mj}c_{nk}^e \,,\\ 
\lambda'_{ijk} = \lambda'_{lmn}c_{li}^Qc_{mj}c_{nk}^D\,,~~ 
\lambda_{ij}^E = \lambda_{lmn} c_{li}c_{mj}c_{n4}^e \,, ~~
y_i^E = 2\lambda_{lmn} c_{li}c_{m4}c_{n4}^e \,,~~ 
\lambda_{ij}^Q = \lambda'_{lmn} c_{l4}^Qc_{mi}c_{nj}^D \,,\\ 
y_i^{Q\prime}  = \lambda'_{lmn} c_{l4}^Qc_{m4}c_{ni}^D \,, 
\lambda_{ij}^D = \lambda'_{lmn} c_{li}^Qc_{mj}c_{n4}^D \,,~~ 
y_i^D = \lambda'_{lmn} c_{li}^Qc_{m4}c_{n4}^D \,,~~ 
\lambda_i^{QD} = \lambda'_{lmn} c_{l4}^Qc_{mi}c_{n4}^D \,,\\ 
y^{QD} = \lambda'_{lmn} c_{l4}^Qc_{m4}c_{n4}^D \,,~~ 
y_i^U  = y_{mn} c_{mi}^Qc_{n4}^U \,,~~  
y_i^Q  = y_{mn} c_{m4}^Qc_{ni}^U \,,~~ 
y^{QU} = y_{mn} c_{m4}^Qc_{n4}^U \,,\\ 
\lambda'_i = y'_m c_{mi} \,, ~~
y = y'_m c_{m4} \,,~~ 
\tilde{\lambda}_i = \tilde{y}_m c_{mi}^D \,, ~~ 
\tilde{\lambda} = \tilde{y}_m c_{m4}^D \,,\\
\tilde{\lambda}_{ij} = \tilde{y}_{mn} c_{mi}^ec_{nj}^U \,,~~ 
\tilde{\lambda}_i^U = \tilde{y}_{mn} c_{mi}^ec_{n4}^U \,,~~ 
\tilde{\lambda}_i^E = \tilde{y}_{mn} c_{m4}^ec_{ni}^U \,,~~ 
\tilde{\lambda}^{EU}= \tilde{y}_{mn} c_{m4}^ec_{n4}^U \,. 
\end{array}
\end{equation} 

From the superpotential (\ref{4}), we see that because of Dirac mass 
terms of up-type Higgs and the four generations, one of the four 
generation, namely the fourth generation 
($H_d$, $E_4^c$, $Q_4$, $U_4^c$, $D_4^c$), is always heavy which is 
identified as the one containing the down-type Higgs.  The fourth 
generation neutrino together with the "neutrino" in $H_u$ consists of 
Higgsinos.  After the mass terms, the next five terms in Eq. (\ref{4}) 
are ordinary Yukawa interactions and trilinear lepton number (R-parity) 
violating terms, where $i,~j,~k$ stand for three light generations.  The 
other terms in (\ref{4}) are new which involve extra generations.  
Many of these new terms violate lepton numbers.  Note that by taking 
$\mu^{e,Q,U,D}\gg \mu\sim {\mathcal O}$(TeV), we obtain MSSM as a 
low energy effective theory.  

\subsection{SUSY breaking}  

Soft SUSY breaking mass terms should be included into the Lagrangian.  In 
addition to gaugino masses, they include mass-squared terms of scalars 
and $B\mu$-type terms corresponding to those $\mu$-terms in 
superpotential (\ref{1}), 
\begin{equation}
\label{7} 
\begin{array}{lll}
-{\mathcal L}&\supset&
                   M^2\tilde{L}_m^{\dag}\tilde{L}_m + M_h^2h_u^{\dag}h_u 
                      + M_E^2\tilde{E}_m^{c\dag}\tilde{E}_m^c 
                      + M_Q^2\tilde{Q}_m^{\dag}\tilde{Q}_m 
                      + M_U^2\tilde{U}_m^{c\dag}\tilde{U}_m^c 
                      + M_D^2\tilde{D}_m^{c\dag}\tilde{D}_m^c \\
            &        &+ M_{EH}^2\tilde{E}_H^{c*}\tilde{E}_H^c 
                      + M_{QH}^2\tilde{Q}_H^{\dag}\tilde{Q}_H 
                      + M_{UH}^2\tilde{U}_H^{c*}\tilde{U}_H^c 
                      + M_{DH}^2\tilde{D}_H^{c*}\tilde{D}_H^c \\
           &&+(B\mu_m\tilde{L}_mh_u+B^e\mu_m^e\tilde{E}_m^c\tilde{E}_H^c 
                        +B^Q\mu_m^Q\tilde{Q}_m\tilde{Q}_H
                        +B^U\mu_m^U\tilde{U}_m^c\tilde{U}_H^c 
                        +B^D\mu_m^D\tilde{D}_m^c\tilde{D}_H^c + h.c.)\,,
\end{array}
\end{equation}
where tildes stand for scalars.  We have assumed universality of the 
mass-squared terms and the alignment of the $B$ terms, namely the mass 
parameters $B^{e,Q,U,D}$ do not depend on the sub-script $m$.  In terms 
of three light generations of Eq. (\ref{4}), universality of these soft 
mass terms is easily seen, 
\begin{equation}
\label{8} 
\begin{array}{lll}
-{\mathcal L}&\supset& M^2\tilde{L}_i^{\dag}\tilde{L}_i+M^2h_d^{\dag}h_d  
            + M_h^2h_u^{\dag}h_u + M_E^2\tilde{E}_m^{c\dag}\tilde{E}_m^c 
                      + M_Q^2\tilde{Q}_m^{\dag}\tilde{Q}_m 
                      + M_U^2\tilde{U}_m^{c\dag}\tilde{U}_m^c 
                      + M_D^2\tilde{D}_m^{c\dag}\tilde{D}_m^c \\
             &        &+ M_{EH}^2\tilde{E}_H^{c*}\tilde{E}_H^c 
                      + M_{QH}^2\tilde{Q}_H^{\dag}\tilde{Q}_H 
                      + M_{UH}^2\tilde{U}_H^{c*}\tilde{U}_H^c 
                      + M_{DH}^2\tilde{D}_H^{c*}\tilde{D}_H^c \\
             &       &+(B\mu h_dh_u+B^e\mu^e\tilde{E}_4^c\tilde{E}_H^c 
                        +B^Q\mu^Q\tilde{Q}_4\tilde{Q}_H
                        +B^U\mu^U\tilde{U}_4^c\tilde{U}_H^c 
                        +B^D\mu^D\tilde{D}_4^c\tilde{D}_H^c + h.c.)\,.
\end{array}
\end{equation}
Numerically soft masses $M$'s, $B$'s and gaugino masses are assumed to 
be $\cal O$(100) GeV.  

Soft trilinear terms corresponding to Eq. (\ref{1}) are 
\begin{equation}
\label{9}
\begin{array}{lll}
{\mathcal L}&\supset&
               \bar{\lambda}_{lmn} \tilde{L}_l \tilde{L}_m \tilde{E}_n^c 
               +\bar{\lambda}'_{lmn} \tilde{Q}_l\tilde{L}_m\tilde{D}_n^c 
                +\bar{y}_{mn} \tilde{Q}_m h_u \tilde{U}_n^c \\
            & & +\bar{y}'_m \tilde{Q}_H \tilde{L}_m \tilde{U}_H^c 
               +\bar{\tilde{y}}_m\tilde{E}_H^c\tilde{D}_m^c\tilde{U}_H^c 
            +\bar{\tilde{y}}_{mn}\tilde{E}_m^c\tilde{D}_H^c\tilde{U}_n^c 
                + h.c. \,,
\end{array} 
\end{equation} 
where the following coupling alignment will be assumed, 
\begin{equation} 
\label{10} 
\bar{\lambda}_{lmn}^{(\prime)} = \lambda_{lmn}^{(\prime)}m_0\,,~~
\bar{y}_{mn} = y_{mn}m_0\,,~~  \bar{y}'_m = y'_mm_0 
\end{equation}
with $m_0$ being the order of soft masses $\sim{\cal O}(100)$ GeV.  

\subsection{EWSB}  

Let us look at gauge symmetry breaking.  
From the Lagrangian, the scalar potential can be written down 
straightforwardly.  To get EWSB, one needs a negative determinant of 
the Higgs mass-squared matrix, namely 
\begin{equation}
\label{10a}
\left(M^2  +\mu^2\right) \left(M_h^2+\mu^2\right) < |B\mu|^2\,
\end{equation}
with the ordinary condition $M^2+M_h^2+2\mu^2+2B\mu>0$.  This 
requirement can be realized when the renormalization group is considered.  
$M_h^2$ will become negative at the weak scale, due to the large top 
quark Yukawa coupling.  Therefore, everything of EWSB here will be the 
same as that in MSSM.  The MSSM analysis of EWSB applies here.  EWSB 
in this model occurs at the weak scale.   

In addition to $B\mu$, other $B\mu$-terms in Eqs. (\ref{7}) and (\ref{8}) 
might complicate the gauge symmetry breaking analysis.  The experience 
from MSSM shows that if a $B\mu$-term is large enough, gauge symmetry 
breaking always occurs.  The new $B^{Q,U,D}\mu^{Q,U,D}$-terms could 
result in color symmetry breaking and the $B^e\mu^e$-term could cause 
purely U(1)$_Y$ symmetry breaking.  Such unwanted gauge symmetry breaking 
should be avoided.  To be concrete, besides Eq. (\ref{10a}), correct EWSB 
also requires 
\begin{equation}
\label{13} 
\left(M_X^2+\mu^{X2}\right)\left(M_{XH}^2+\mu^{X2}\right) >  
|B^X\mu^X|^2~~~{\rm for}~~~ X=e,Q,U,D\,.   
\end{equation}
Then the remaining analysis of EWSB is identical to that of MSSM with 
same Higgs and Higgsino spectra.  Eq. (\ref{13}) can be satisfied easily.  
Careful thinking of EWSB conditions Eqs. (\ref{10a}) and (\ref{13}), we 
see that if $\mu<\mu^X$, EWSB occurs naturally.  This point will be 
discussed later.  The fact that pure U(1)$_Y$ breaking does not occur 
can be simply due to a large enough $\mu^e$ compared to $\mu$.  

In the scalar potential, quartic terms are 
determined by SUSY gauge interactions.  
From the point of view of quartic terms, the larger the coefficients of 
certain quartic terms, the more difficult the gauge symmetry breaking 
is.  Therefore EWSB is easier to be obtained compared to that of the 
color symmetry.  Traditional SU(2)$_L\times$U(1)$_Y$ breaking via Higgs 
doublet fields contains the quartic term 
\begin{equation} 
\label{11} 
V \supset \frac{g^2+g^{\prime 2}}{8} 
          \left(h_u^\dag h_u-h_d^\dag h_d\right)^2 + ......\,. 
\end{equation} 
Whereas for purely U(1)$_Y$ symmetry breaking, the relevant quartic term 
is 
\begin{equation}
\label{12}
V \supset \frac{g^{\prime 2}}{2} \left(\tilde{E}_H^{c*}\tilde{E}_H^c 
                     -\tilde{E}_4^{c*}\tilde{E}_4^c\right)^2 + ......\,. 
\end{equation} 
Because $g^{\prime 2}/2$ is comparable to $(g^2+g^{\prime 2})/8$ 
numerically at the weak scale, pure U(1)$_Y$ symmetry breaking is not 
really favored.

\subsection{Fermion spectra}

We write down relevant matter superfields in SU(2)$_L$ components 
explicitly, 
$L_i=\displaystyle\left(
\begin{array}{c}
L_i^0 \\ L_i^- 
\end{array}
\right)$, 
$Q_m=\displaystyle\left(
\begin{array}{c}
Q_m^t \\ Q_m^b 
\end{array}
\right)$ and 
$Q_H=\displaystyle\left(
\begin{array}{c}
Q_H^b \\ Q_H^t 
\end{array}
\right)$.  Because fourth generation doublet leptons have been taken as 
Higgsinos, and EWSB is the same as that in MSSM, gauginos and Higgsinos 
are identical to those in MSSM.  In the following we first look at the 
lepton spectrum.  

As we have seen, the first three generation sneutrinos do not obtain 
vacuum expectation values (VEVs).  SU(2)$_L$ doublet leptons, therefore, 
do not mix with the down-type Higgsino and chargino.  Lepton masses are 
due to ordinary Yukawa couplings and Yukawa couplings between SU(2)$_L$ 
doublet leptons with the fourth generation singlet lepton, as well as 
the $\mu^e$ term, 
\begin{equation}
\label{14}
{\mathcal L} \supset -\left(e_i^-, e_H^c\right) {\mathcal M}^l 
\left(
\begin{array}{c}
e_j^c \\ e_4^c 
\end{array}
\right) \,,
\end{equation}
where small letters denote fermionic components, the $4\times4$ charged 
lepton mass matrix is given as 
\begin{equation}
\label{15}
{\mathcal M}^l = \left(
\begin{array}{cc}
m_{ij}^l & m_{i4}^l \\
0        & \mu^e    \\[3mm] 
\end{array}
\right) \,, 
\end{equation}
where $m_{ij}^l\equiv y_{ij}^l\displaystyle\frac{v}{\sqrt{2}}\cos\beta$, 
$m_{i4}^l\equiv y_i^E\displaystyle\frac{v}{\sqrt{2}}\cos\beta$.  
Consider typically the 3rd and 4th generation case, namely that of $i,j$ 
being 3, taking $\displaystyle\frac{|m_{33}^l||m_{34}^l|}
{|\mu^e|^2-|m_{33}^l|^2+|m_{34}^l|^2}\ll 1$, lepton masses are obtained, 

\begin{equation}
\label{16}
\begin{array}{lll}
m_\tau & \simeq & \sqrt{|m_{33}^l|^2-|m_{33}^l||m_{34}^l|}\,,\\[3mm] 
M_l    & \simeq & |\mu^e|\,.
\end{array}
\end{equation}
The unitary matrix diagonalizing ${\cal M}^l{\cal M}^{l\dag}$ is then 
\begin{equation}
\label{17}
\left(
\begin{array}{cc}
1 & \displaystyle -\left(1+\frac{|m_{33}^l |}{|m_{34}^l |}\right)
\frac{m_{34}^{l*} }{\mu^{e*}} \\[3mm]
\displaystyle \frac{m_{34}^l}{\mu^e} & 1 
\end{array}
\right) +{\mathcal O}\displaystyle\left(\frac{m_\tau}{\mu^e}\right)^2\,.  
\end{equation}
This implies that there is an 
${\cal O}\left(\displaystyle\frac{m_\tau}{\mu^e}\right)^2$ unitarity 
deviation among the three generation leptons.  

For the down-type quark spectrum, introduction of additional two 
generations makes the full down quark mass matrix 
${\cal M}^d$ being $5\times 5$ one, 
\begin{equation}
\label{18}
{\mathcal L} \supset -\left(q_i^b, q_4^b, d_H^c\right) {\mathcal M}^d 
\left(
\begin{array}{c}
d_j^c \\ d_4^c \\ q_H^t 
\end{array}
\right) \,,
\end{equation}
and 
\begin{equation}
\label{19}
{\mathcal M}^d = \left(
\begin{array}{ccc}
m_{ij}^d & m_{i4}^d & 0      \\
m_{4j}^d & m_{44}^d & -\mu^Q \\
0        & -\mu^D   & 0
\end{array}
\right) \,, 
\end{equation}
where 
$m_{ij}^d\equiv y_{ij}^d\displaystyle\frac{v}{\sqrt{2}}\cos\beta$, 
$m_{i4}^d\equiv y_i^D\displaystyle\frac{v}{\sqrt{2}}\cos\beta$, 
$m_{4i}^d\equiv y_i^{Q\prime}\displaystyle\frac{v}{\sqrt{2}}\cos\beta$, 
$m_{44}^d\equiv y^{QD}\displaystyle\frac{v}{\sqrt{2}}\cos\beta$.  The 
$3\times 3$ sub-matrix $m_{ij}^d$ is the ordinary down quark mass matrix 
which is now not necessarily unitary.  Focusing on its unitarity 
deviation due to extra generations, we consider the sub-mass-matrix of 
the 3rd generation and extra generations, that is ${\cal M}^d$ with $i$ 
and $j$ being 3.  It is natural to take 
$|\mu^Q|\sim|\mu^D|\gg|m_{4j}^d|,~|m_{44}^d|,~|m_{ij}^d|,~|m_{i4}^d|$.  
To the first order of $m_{mn}^d/\mu^D$ with $m$ and $n$ being 3 and 4, 
the absolute mass eigenvalues are then $|\mu^Q|$, $|\mu^D|$ and 
$|m_{33}^d|$, respectively.  The mass matrix is diagonalized by unitary 
matrices $U^d$ and $V^d$, 
\begin{equation}
\label{20}
U^{d\dag}{\mathcal M}^d V^d = \left(
\begin{array}{ccc}
m_b & 0     & 0     \\
0   & \mu^Q & 0     \\
0   & 0     & \mu^D 
\end{array}
\right) 
\end{equation}
where 
\begin{equation}
\label{21}
U^d = \left(
\begin{array}{ccc}
1 & 0 & \displaystyle -\frac{m_{34}^d}{\mu^D} \\[3mm] 
0 & 1 & \displaystyle -\frac{|m_{34}^d|^2}{\mu^D m_{44}^{d*}} \\[3mm] 
\displaystyle-\frac{m_{34}^{d*}}{\mu^{D*}}&
\displaystyle\frac{\mu^D m_{44}^{d*}}{|\mu^D|^2-|\mu^Q|^2} & 1 
\end{array}
\right) +{\mathcal O}\displaystyle\left(\frac{m_b}{\mu^D}\right)^2\,,  
\end{equation} 
and 
\begin{equation}
\label{22}
V^d = \left(
\begin{array}{ccc}
1 & \displaystyle -\frac{m_{43}^{d*}}{\mu^{Q*}}  & 0 \\[3mm] 
0 & \displaystyle \frac{|m_{43}^d|^2}{\mu^{Q*} m_{44}^d}  & 1 \\[3mm] 
\displaystyle \frac{m_{43}^d}{\mu^Q}& 1 & 
\displaystyle\frac{\mu^{Q*} m_{44}^d}{|\mu^Q|^2-|\mu^D|^2}  
\end{array}
\right) +{\mathcal O}\displaystyle\left(\frac{m_b}{\mu^D}\right)^2\,.  
\end{equation} 
Therefore there is generally an 
${\cal O}\left(\displaystyle\frac{m_b}{\mu^D}\right)^2$ 
contribution to unitarity deviation of the Cabbibo-Kobayashi-Maskawa 
(CKM) matrix.  

Similarly, the up-type quark mass matrix ${\cal M}^u$ is given in the 
following 
\begin{equation}
\label{23}
{\mathcal L} \supset -\left(q_i^t, q_4^t, u_H^c\right) {\mathcal M}^u 
\left(
\begin{array}{c}
u_j^c \\ u_4^c \\ q_H^b 
\end{array}
\right) \,,
\end{equation}
and 
\begin{equation}
\label{24}
{\mathcal M}^u = \left(
\begin{array}{ccc}
m_{ij}^u & m_{i4}^u & 0      \\
m_{4j}^u & m_{44}^u & \mu^Q \\
0        & \mu^U    & -m_H^u
\end{array}
\right) \,, 
\end{equation}
where $m_{ij}^u\equiv -y_{ij}^u\displaystyle\frac{v}{\sqrt{2}}\sin\beta$, 
$m_{i4}^u\equiv -y_i^U\displaystyle\frac{v}{\sqrt{2}}\sin\beta$, 
$m_{4j}^u\equiv -y_j^Q\displaystyle\frac{v}{\sqrt{2}}\sin\beta$, 
$m_{44}^u\equiv -y^{QU}\displaystyle\frac{v}{\sqrt{2}}\sin\beta$, and 
$m_H^u\equiv y\displaystyle\frac{v}{\sqrt{2}}\cos\beta$.  Taking that 
$|\mu^Q|\sim|\mu^U|\gg|m_{4j}^u|,~|m_{44}^u|,~|m_{ij}^u|,~|m_{i4}^u|,~
|m_H^u|$, the absolute mass eigenvalues are $|\mu^Q|$, $|\mu^U|$, and 
$|m_{33}^u|$, respectively.  The diagonalizing matrices to Eq. (\ref{24}) 
are $U^u$ and $V^u$, 
\begin{equation}
\label{25}
U^{u\dag}{\mathcal M}^u V^u = \left(
\begin{array}{ccc}
m_t & 0     & 0     \\
0   & \mu^Q & 0     \\
0   & 0     & \mu^U 
\end{array}
\right) 
\end{equation}
where 
\begin{equation}
\label{26}
U^u = \left(
\begin{array}{ccc}
1 & 0 & \displaystyle \frac{m_{34}^u}{\mu^U} \\[3mm] 
0 & 1 & \displaystyle 
\frac{|m_H^u|^2+|m_{34}^u|^2}{\mu^{Q*}m_H^u-\mu^U m_{44}^{u*}} \\[3mm] 
\displaystyle-\frac{m_{34}^{u*}}{\mu^{U*}}&
\displaystyle
\frac{\mu^U m_{44}^{u*}-m_H^u\mu^{Q*}}{|\mu^Q|^2-|\mu^U|^2-|m_H^u|^2}& 1 
\end{array}
\right) +{\mathcal O}\displaystyle\left(\frac{m_t}{\mu^U}\right)^2\,,  
\end{equation} 
and 
\begin{equation}
\label{27}
V^u = \left(
\begin{array}{ccc}
1 & \displaystyle \frac{m_{43}^{u*}}{\mu^{Q*}}  & 0 \\[3mm] 
0 & \displaystyle 
\frac{|m_H^u|^2+|m_{43}^u|^2}{\mu^Um_H^{u*}-\mu^{Q*} m_{44}^u}&1\\[3mm] 
\displaystyle-\frac{m_{43}^u}{\mu^Q}& 1 & 
\displaystyle
\frac{\mu^{Q*} m_{44}^u-m_H^{u*}\mu^U}{|\mu^U|^2-|\mu^Q|^2-|m_H^u|^2}  
\end{array}
\right) +{\mathcal O}\displaystyle\left(\frac{m_t}{\mu^U}\right)^2\,.  
\end{equation} 
The contribution to unitarity deviation of the CKM matrix can be as 
large as $\left(m_t/\mu^U\right)^2$.

\subsection{Boson spectra}  

Boson masses are more complicated than the fermion case because of soft 
terms as well as SUSY kinetic terms.  Higgs bosons, and therefore gauge 
bosons, in this model have exactly the same spectra as those in MSSM.  
So we will only consider slepton and squark masses.  

SUSY kinetic terms contribute sfermion masses in the same manner like 
that in MSSM, 
\begin{equation}
\label{28}
V \supset m_Z^2\cos 2\beta \left(I_3-\sin^2\theta_W Q\right)f^*f 
\end{equation} 
where $\theta_W$ is the Weinberg angle, and the field $f$ denotes 
$\tilde{L}_i,~ \tilde{Q}_m,~ \tilde{Q}_H,~ \tilde{E}_{Ri},~ \tilde{E}_4,
~\tilde{E}_H,~ \tilde{U}_m^c,~ \tilde{D}_m^c,~ \tilde{D}_H^c,~ 
\tilde{U}_H^c$.  

Trilinear terms in Eqs. (\ref{9}) and (\ref{10}) contribute 
\begin{equation}
\label{29}
{\mathcal L} \supset -m_0\left(m_{im}^l \tilde{L}_i^-\tilde{E}_m^c
+m_{mn}^d\tilde{Q}_m^b\tilde{D}_n^c+m_{mn}^u\tilde{Q}_m^t\tilde{U}_n^c
+m_H^u\tilde{Q}_H^b\tilde{U}_H^c + h.c.\right)\,.
\end{equation}
The superpotential which involves $\mu$ terms and Yukawa interactions 
contributes 
\begin{equation}
\label{30}
\begin{array}{lll}
{\mathcal L} & \supset & -[|\mu|^2\left(h_d^\dag h_d+h_u^\dag h_u\right) 
+ |\mu^e|^2 \left(\tilde{E}_4^{c*} \tilde{E}_4^c 
+ \tilde{E}_H^{c*}\tilde{E}_H^c \right) 
+ |\mu^Q|^2 \left( \tilde{Q}_4^\dag \tilde{Q}_4 
+ \tilde{Q}_H^\dag \tilde{Q}_H \right) \\ 
& & + |\mu^U|^2 \left( \tilde{U}_4^{c*} \tilde{U}_4^c 
+ \tilde{U}_H^{c*} \tilde{U}_H^c \right) 
+ |\mu^D|^2 \left( \tilde{D}_4^{c*} \tilde{D}_4^c 
+ \tilde{D}_H^{c*} \tilde{D}_H^c \right) 
+ |m_H^u|^2 \left( \tilde{U}_H^{c*} \tilde{U}_H^c 
+ \tilde{Q}_H^{b*} \tilde{Q}_H^b \right) \\ 
& & + \tilde{E}^{c*} m^{l\dag}m^l\tilde{E}^c 
+\tilde{D}^{c*} m^{d\dag}m^d\tilde{D}^c 
+ \tilde{U}^{c*} m^{u\dag}m^u\tilde{U}^c \\
& & + \tilde{L}m^lm^{l\dag}\tilde{L}^*
+ \tilde{Q}^b m^dm^{d\dag} \tilde{Q}^{b*}
+ \tilde{Q}^t m^um^{u\dag} \tilde{Q}^{t*} \\ 
& & -(\mu\tan\beta\tilde{L} m^l \tilde{E}^c  
+ \mu\tan\beta\tilde{Q}^b m^d \tilde{D}^c 
+ \mu\cot\beta\tilde{Q}^t m^u \tilde{U}^c 
+ \mu\tan\beta m_H^u \tilde{Q}_H^d \tilde{U}_H^c \\
& & + \mu^{e*}m^l_{i4} \tilde{L}_i^- \tilde{E}_H^{c*} 
+ \mu^{D*}m^d_{m4} \tilde{Q}_m^b \tilde{D}_H^{c*} 
+ \mu^{U*}m^u_{m4} \tilde{Q}_m^t \tilde{U}_H^{c*} 
+ \mu^{Q*}m_H^u \tilde{Q}_4^{t*} \tilde{U}_H^{c*} \\
& & - \mu^{Q*}m^d_{4m} \tilde{Q}_H^{t*} \tilde{D}^c_m 
+ \mu^{Q*} m^u_{4m} \tilde{Q}_H^{b*} \tilde{U}^c_m 
+\mu^{U*}m_H^u \tilde{Q}_H^b \tilde{U}^{c*}_4 + h.c. ) ] \,,
\end{array}
\end{equation}
where $m^{l,d,u}$ are matrices with elements defined before.  

Together with $B\mu$ terms given in Eq. (\ref{8}), full mass-squared 
matrices of sfermions are obtained.  In the Appendix, the charged 
slepton, up squark and down squark mass-squared matrices will be 
given explicitly.

\section{Phenomenological constraints}  

The direct experimental search of extra generation particles at LEP 
requires that they should be heavier than 100 GeV, and direct search of 
extra generation quarks at Tevatron requires they are heavier than 270 
GeV \cite{pdg}.  This result can be simply satisfied if $\mu^X$'s are 
larger than 100 GeV or 270 GeV.  Note that we do not have extra 
neutrinos which, in this model, consist of Higgsinos.  

The electroweak precision measurement generally has constraints on extra 
matters \cite{stu}.  
Current constraints are \cite{pdg} 
\begin{equation}
\label{10b}
\begin{array}{lll}
S & = & -0.10 \pm 0.10 \,,\\ 
T & = & -0.08 \pm 0.11 \,,\\
U & = &  0.15 \pm 0.11 \,.
\end{array}
\end{equation}
For one extra chiral generation, oblique parameters 
$S$, $T$, $U$ still allow the existence of the 4th generation \cite{stu} 
provided that there is certain mass splitting in extra SU(2)$_L$ 
doublets.  In our case, the vector-like generation contributes to the 
parameters in the way of $1/\mu^{X2}$ as expected from the decoupling 
theorem \cite{decoupling}.  Typically, 
\begin{equation}
\label{10c}
S \simeq T \simeq \left(\frac{m_t}{\mu^X}\right)^2 \,. 
\end{equation}
The effect of the extra generation can be 
small enough $\leq\left(m_t/\mu^X\right)^2\simeq (1-10)\%$ if we take 
$\mu^X\simeq 500$ GeV $-1$ TeV.  

Important constraints come from the unitarity of the $3\times 3$ CKM 
quark mixing matrix of three chiral generations \cite{pdg}.  This 
unitarity is consistent with current data within experimental errors.  
In this model, extra generations mix with ordinary three chiral 
generations which necessarily break the unitarity of the CKM mixing 
matrix.  As we have observed following Eqs. (\ref{19}) and (\ref{24}), 
unitarity violation is about $(m_{i4}^{d(u)}/\mu^{D(U)})^2$.  This 
$\mu^X$ dependence is generally expected in the case of extra 
vector-like generations.  Hierarchical or small mixing masses 
$m_{i4}^{d(u)}$ can easily make the CKM matrix approximately unitary 
within errors.  For an example, $(m_{14}^{d(u)}/\mu^X)^2\leq 10^{-3}$.  
Assuming only the third generation mixes with extra generations, the 
constraint is still loose, $(m_{34}^{d(u)}/\mu^{D(U)})^2\leq 0.39$.  The 
quantity $m_{34}^{d(u)}$ is at most about $m_t$.  This gives that 
the parameter $\mu^{D,U}\geq 280$ GeV.  

From Eqs. (\ref{21}), (\ref{22}), (\ref{26}) and (\ref{27}), it can be 
seen that there are new phases in fermion mixing matrices.  However, 
these new matrix elements are of order of $\left(m_t/\mu^X\right)^2$ at 
most.  So new CP violation effects are generally suppressed.  

One of the characteristic properties of the superpotential in this model 
is that there are many $\mu$-parameters.  In the following analysis, we 
prefer to take all the $\mu$-parameters approximately equal.  They are 
expected having common origin.  On the other hand, by taking 
$|\mu|\ll |\mu^X|$, such as $|\mu|=100$ GeV and $|\mu^X|=500$ GeV, 
this model will have MSSM as its low energy 
effective theory.  In order to make this model distinct, considering 
above discussed constraints, we will take $|\mu|$ close to $|\mu^X|$, 
and 
\begin{equation}
\label{31}
|\mu^X| \sim 500 ~{\rm GeV}\;.
\end{equation} 
We bear in mind that for correct EWSB, it maybe necessary to require 
that $|\mu|$ is a kind of smaller than $|\mu^e|$, and $|\mu|$ cannot be 
too large.  
Of course, values of soft masses and $B\mu$'s are also important to 
EWSB.  They can affect the choice of values of $\mu$-parameters.  

In the following, we give a numerical illustration.  $\mu$ parameters 
can have the following values, 
\begin{equation}
\label{335}
\mu = 300~{\rm GeV}\,,~~\mu^e = 400~{\rm GeV}\,,~~
\mu^{Q,U,D} = 500 {\rm GeV}\,.  
\end{equation}
Correct EWSB happens if we take 
\begin{equation}
\label{336}
\begin{array}{lll}
B\mu & = & -(260~{\rm GeV})^2\,,~~M^2 = (219~{\rm GeV})^2\,,~~
M_h^2 = -(243~{\rm GeV})^2\,,\\
B^X\mu^X & = & -(200~{\rm GeV})^2\,,~~
M_X^2 = M_{XH}^2 = (200~{\rm GeV})^2\,.  
\end{array}
\end{equation}
Note that we have taken $M_h^2$ negative.  This is expected due to the 
large top quark Yukawa coupling \cite{susy}.  It is straightforward to 
see that this set of parameters results in consistent EWSB, and it fixes 
that $\tan\beta = 2$ and results in the Higgs spectrum, 
\begin{equation}
\label{337}
m_h^2 \simeq (124~{\rm GeV})^2\,,~~m_{H^0}^2 \simeq (420~{\rm GeV})^2\,,~~
m_A^2 \simeq (413~{\rm GeV})^2\,,~~m_{H^\pm}^2 \simeq (424~{\rm GeV})^2\,,
\end{equation}
where the quantum correction to $m_h^2$ has been included.

\section{LHC phenomenology}

This model can be tested at LHC.  In addition to the (super) particle 
content of MSSM, it predicts following vector-like particles: one lepton 
singlet ($E_4^c$, $E_H^c$), one quark doublet ($Q_4$, $Q_H$), one 
up-type quark singlet ($U_4^c$, $U_H^c$) and one down-type quark singlet 
($D_4^c$, $D_H^c$), but there is no extra doublet leptons which are 
already identified as the Higgs doublets.  

Let us look at fermions.  Because the effect of EWSB is much smaller 
than $\mu$-parameters, these new Weyl fermions form several Dirac 
fermions with masses $|\mu^X| \sim 500$ GeV, 
\begin{equation}
\label{32}
\Psi_e \equiv \left(\begin{array}{c}
e_H^c\\
\bar{e}_4^c
\end{array}
\right)\;,
\Psi_Q \equiv \left(\begin{array}{c}
q_4\\
\bar{q}_H
\end{array}
\right)\;,
\Psi_u \equiv \left(\begin{array}{c}
u_4^c\\
u_H
\end{array}
\right)\;,
\Psi_d \equiv \left(\begin{array}{c}
d_4^c\\
d_H
\end{array}
\right) \; .  
\end{equation}
Note that in this case, mass splitting in the SU(2)$_L$ doublet $q_4$ or 
$q_H$ is also negligible.  Considering gauge kinetic terms, the 
Lagrangian describing these pseudo-Dirac fermions can be written as 
\begin{equation}
\label{33}
{\mathcal L} \supset \sum_X \left( \bar{\Psi}_X \gamma_\mu D^\mu \Psi_X + 
\mu^X \bar{\Psi}_X \Psi_X \right) \; , 
\end{equation} 
where the covariant derivative is self-evident.  

Decay signals of these new particles can be easily identified.  From 
trilinear Yukawa interactions given in Eq. (\ref{4}), it is seen that 
they decay into SM first three generation matters.  $\Psi_e$ can decay 
into $e_i$ and a neutral Higgs, the decay rate is  
\begin{equation}
\label{34}
\Gamma(\Psi_e \to e_i h^0) \simeq \frac{1}{16\pi}|y^E_i|^2 |\mu^e|
\left(1-\frac{m_h^2}{|\mu^e|^2}\right)^2 \;.
\end{equation} 
Alternative to the Higgs, scalar neutrinos can be also the decay 
product which is expected at least heavier than the lighter neutral 
Higgs.  Similarly, new quarks $\Psi_u$ and $\Psi_d$ decay into 
ordinary quarks and the Higgs, 
\begin{equation}
\label{35}
\Psi_u \to u \; (c\;, t) \; \; h^0 \;,~~~~~
\Psi_d \to d \; (s\;, b) \; \; h^0 
\end{equation} 
with decay rates 
\begin{equation}
\label{36}
\Gamma(\Psi_{u(d)} \to u_i (d_i) \; \; h^0) \simeq 
\frac{1}{16\pi}|y^{4(D)}_i|^2 |\mu^{U(D)}|
\left(1-\frac{m_h^2}{|\mu^{U(D)}|^2}\right)^2  \;.
\end{equation} 
And decays of new quarks $q_4$ and $q_H$ have following results, 
\begin{equation}
\label{37}
\Gamma(\Psi_q^{d(u)} \to d_i^c (u_i^c) \; \; h^0) \simeq 
\frac{1}{16\pi}|y^{Q'(Q)}_i|^2 |\mu^Q|
\left(1-\frac{m_h^2}{|\mu^Q|^2}\right)^2  \;.
\end{equation} 
Taking relevant Yukawa coefficients $y_i$'s $\sim~10^{-1}-10^{-2}$, 
decay rates in Eqs. (\ref{34})-(\ref{37}) are $\Gamma \sim 5-500$ MeV.  

Taking EWSB into consideration, $\Psi_X$ mixes with SM fermions.  The 
$5 \times 5$ generalized CKM matrix is derived from Eqs. (\ref{21}) and 
(\ref{26}), 
\begin{equation}
\label{38}
V \equiv U^{u\dag} U^d \simeq  \left(
\begin{array}{ccc}
1 & 0 & \displaystyle -\frac{m_{34}^d}{\mu^D}-\frac{m_{34}^u}{\mu^U} \\[3mm]
0 & 1 & \displaystyle \frac{|m_{34}^d|^2}{\mu^D m_{44}^{d*}}
+\frac{\mu^{U*}m_{44}^U-m_H^{u*}\mu^D}{|\mu^Q|^2-|\mu^U|^2-|m_H^u|^2} \\[3mm]
\displaystyle \frac{m_{34}^{u*}}{\mu^{U*}}+\frac{m_{34}^{d*}}{\mu^{D*}}  & 
\displaystyle \frac{|m_H^u|^2+|m_{i4}^u|^2}{\mu^Q m_H^{u*}-\mu^{U*}m_{44}^u} & 1
\end{array}
\right) \,.
\end{equation}  
Note that ${\mathcal O}(m/\mu^X)^2$ terms have been omitted.  We see 
that decays 
$\Psi_d \rightarrow \bar{t}~W^+$ and $\Psi_u \rightarrow \bar{b}~W^-$ 
occur via the SU(2)$_L$ gauge interaction at the level of 
${\mathcal O}(m/\mu^X)$, 
\begin{equation}
\label{39}
\begin{array}{lll}
\Gamma(\Psi_d \rightarrow \bar{t}~W^+) & \simeq & \displaystyle
\frac{G_F m_W^2|\mu^D||V_{35}|^2}{8\sqrt{2}\pi}
\left\{1+\left(\frac{m_W}{\mu^D}\right)^4+\left(\frac{m_t}{\mu^D}\right)^4 
-2\left[\left(\frac{m_W}{\mu^D}\right)^2
+\left(\frac{m_W}{\mu^D}\right)^2\left(\frac{m_t}{\mu^D}\right)^2
\right.\right.
\\[3mm]
& & \displaystyle \left.\left. 
+\left(\frac{m_t}{\mu^D}\right)^2\right]\right\}^{1/2}
\left\{\left[1-\left(\frac{m_t}{\mu^D}\right)^2\right]^2 
+\left(\frac{m_W}{\mu^D}\right)^2\left[1+
\left(\frac{m_t}{\mu^D}\right)^2\right]-2\left(\frac{m_W}{\mu^D}\right)^4
\right\} \; ,\\[3mm]
\Gamma(\Psi_u \rightarrow \bar{b}~W^-) & \simeq & \displaystyle 
\frac{G_F m_W^2|\mu^U||V_{53}|^2}{8\sqrt{2}\pi}
\left[1-\left(\frac{m_W}{\mu^U}\right)^2\right]^2
\left[1+2\left(\frac{m_W}{\mu^U}\right)^2\right] \; ,
\end{array}
\end{equation}
where the phase space factors were given in Refs. \cite{phase}.  
Taking $m/\mu^X \sim 1/3$, these $\Gamma$'s are about 1 GeV.  

All above decays are fast enough that they occur inside detectors.  
With the invariant mass method, decayed new fermions will be 
reconstructed.  

These new quarks can be produced at LHC via gluon fusion processes, 
$g~~g \rightarrow Q_4 Q_H$, $U_4^c U_H^c$, $D_4^c D_H^c$.  The 
production mechanism is essentially the same as that of the top quark 
\cite{bn} with an estimated cross section $\sim$ hundreds $fb$ by taking 
$\mu^X \sim 500$ GeV and $\sqrt{s}=14$ TeV.

For the new lepton, the Drell-Yan process is the main production 
mechanism, $p~~p \rightarrow \gamma~~Z \rightarrow e^c e_H^c$.  The 
cross section is estimated to be few $fb$ which means a few tens events 
in one year \cite{fc}.  Considering the detector efficiency, new lepton 
observation maybe challenging at LHC.  However, once they are produced, 
their decay signals are easy to be identified.

\section{Summary and discussion}

Within the framework of SUSY, we have extended the matter content in a 
way that each Higgs doublets is in a full generation.  Namely in 
addition to ordinary three generations, there is an extra vector-like 
generation, and it is the extra slepton SU(2)$_L$ doublets that are 
taken as two Higgs doublets.  R-parity violating interactions contain 
ordinary Yukawa interactions.  SUSY and gauge symmetry breaking have 
been analyzed.  Fermion and boson spectra have been calculated.  
Phenomenological constraints and relevant LHC physics have been 
discussed.  

Finally, we discuss some aspects of this model.  
We are motivated by trying to 
naturally understand the Higgs.  Within SUSY, Higgs can be considered as 
certain slepton doublets.  Thus they are nothing special compared to 
three ordinary generations, the anomaly cancels within each generation.  
It might be amusing to note that the first generation composes the 
ordinary matter, the second generation provides fermion mixing, the 
third gives CP violation, and the fourth and fifth (the vector one) give 
out EWSB because they contain Higgs doublets.  

Like in traditional R-parity violating models, baryon number 
conservation is required.  While the requirement of any first principle 
global symmetry is a draw back compared to SM, it is possible to 
consider baryon number conservation as a result of the so-called 
discrete gauge symmetry \cite{discrete}.  

R-parity violation implies that neutralinos cannot be DM.  Of course, 
the TeV particle theory does not necessarily provide DM, there are many 
alternative scenarios, like the axion, the sterile neutrino, or even the 
modified gravity.  Nevertheless, the thermally produced weakly 
interacting massive particle (WIMP) is one of the most attractive.  To 
this model, it is still possible to introduce additional matter playing 
the role of WIMP.  We note that a recent DM proposal by Arkani-Hamed 
$et$ $al.$ \cite{newdm} can be directly combined with our model.  Their 
proposal is an effort to explain all recent DM experiments \cite{dmexp}, 
WIMP DM lies in a new sector.  

Neutrino masses can be generated.  This model has lepton number 
violation which contributes neutrino masses \cite{rpv}.  Although the 
original superpotential (\ref{1}) is simple, its expression in terms of 
ordinary three light generations given in Eq. (\ref{4}) is complicated.  
There are many new lepton number violating sources.  
All of them involve the vector-like extra generation which is heavier 
than soft SUSY breaking masses.  Therefore, except for the ordinary 
R-parity violating terms with couplings $\lambda_{ijk}$ and 
$\lambda'_{ijk}$, the new lepton number violating terms are less 
stringent constrained at low energies.  
The full phenomenological 
analysis of lepton number violation is rather involved, and will be 
considered in a separate work.  Furthermore, the see-saw mechanism can 
be introduced to get the fully realistic neutrino mass pattern.  

It seems that we have lost GUT.  
In MSSM, running gauge coupling constants meet together at the energy 
$\sim 3\times 10^{16}$ GeV.  This is regarded as a result of GUT.  By 
adding new matter which is charged under the SM, gauge coupling 
unification would be lost generally.   However, if the new matter 
composes complete representations of GUT, gauge coupling unification is 
still kept at least to the one-loop level \cite{21}.  Compared to MSSM, 
the new matter contents we have added in this 
model are ($E_4^c$, $Q_4$, $U_4^c$, $D_4^c$) 
and ($E_H^c$, $Q_H$, $U_H^c$, $D_H^c$).  They do not compose complete 
representations of GUT.  
Giving up GUT while keeping SUSY sounds 
bizarre.  However, GUT relation of gauge coupling constants maybe 
finally restored after additional matter, that is DM, is included.  This 
SUSY model has already introduced the new matter 
$\bf{5}\oplus\bf{\bar{5}}$ and $\bf{10}\oplus\bf{\bar{10}}$ in SU(5) 
representation.  Gauge couplings still do not reach their Landau poles 
in the GUT energy scale \cite{21,pati}.  

We have ignored the mixing of first two chiral generations with the 
vector generation.  Detailed consideration of such possibly small mixing 
may give interesting observable phenomena \cite{vf}.  

SUSY breaking and its mediation to our sector should be considered 
systematically. This is closely related to EWSB and LHC phenomenology.  

All above discussed aspects deserve further and separate studies.

\begin{acknowledgments}
I would like to thank Xiao-gang He, Shu-fang Su and Da-xin Zhang for 
helpful 
discussions.  This work was supported in part by the National Natural 
Science Foundation of China under nos.90503002 and 10821504.
\end{acknowledgments}

\appendix*

\section{Scalar mass-squared matrices}

In writing sfermion mass matrices, for simplicity and without losing 
generality, we will omit the first two generations.  Neglecting the 
mixing between $(1,2)$ generations and 
$(3,4)$ generations, the first two 
generation sfermion themselves are the same as those in MSSM.  The 
charged slepton mass-squared matrix is $\tilde{\cal M}_l^2$, 
\begin{equation}
\label{a1}
{\mathcal L} \supset \left(\tilde{L}_3^{-*}, \tilde{E}^c_3,\tilde{E}^c_4, 
\tilde{E}_H^{c*}\right) \tilde{\mathcal M}_l^2 
\left(\begin{array}{c}
\tilde{L}_3^- \\ \tilde{E}_3^{c*} \\ \tilde{E}_4^{c*} \\ \tilde{E}_H^c
\end{array}\right)\,, 
\end{equation} 
where $4\times4$ $\tilde{\cal M}_l^2$ is given by the following matrix 
elements, 
\begin{equation}
\label{a2}
\begin{array}{cl}
(\tilde{\mathcal M}_l^2)_{11}& =
M^2+\left(\frac{m_Z^2}{2}-m_W^2\right)\cos2\beta+m_\tau^2+|m^l_{34}|^2\,,\\
(\tilde{\mathcal M}_l^2)_{12}& = (m_0-\mu\tan\beta)m_\tau^*\,, \\
(\tilde{\mathcal M}_l^2)_{13}& = (m_0-\mu\tan\beta)m^{l*}_{34} \,,\\
(\tilde{\mathcal M}_l^2)_{14}& = \mu^e m^{l*}_{34}\,, \\
(\tilde{\mathcal M}_l^2)_{21}& = (m_0-\mu\tan\beta)m_\tau\,, \\
(\tilde{\mathcal M}_l^2)_{22}& = 
M_E^2-(m_Z^2-m_W^2)\cos 2\beta+m_\tau^2+|m^l_{43}|^2\,, \\
(\tilde{\mathcal M}_l^2)_{23}& = m_\tau m^{l*}_{34}+m_{43}m^{l*}_{44}\,,\\
(\tilde{\mathcal M}_l^2)_{24}& = 0\,, \\
(\tilde{\mathcal M}_l^2)_{31}& = (m_0-\mu\tan\beta)m^l_{34}\,, \\
(\tilde{\mathcal M}_l^2)_{32}& = m_\tau^* m^l_{34}+m_{43}^{l*}m^l_{44}\,,\\
(\tilde{\mathcal M}_l^2)_{33}& = 
|\mu^e|^2+M_E^2-(m_Z^2-m_W^2)\cos 2\beta+|m^l_{44}|^2 +|m^l_{34}|^2\,, \\ 
(\tilde{\mathcal M}_l^2)_{34}& = B^e\mu^e\,, \\ 
(\tilde{\mathcal M}_l^2)_{41}& = \mu^{e*} m^l_{34}\,,\\ 
(\tilde{\mathcal M}_l^2)_{42}& = 0\,, \\
(\tilde{\mathcal M}_l^2)_{43}& = B^{e*}\mu^{e*}\,,\\
(\tilde{\mathcal M}_l^2)_{44}& = 
|\mu^e|^2 + M_{EH}^2 + (m_Z^2-m_W^2)\cos 2\beta\,.  
\end{array}
\end{equation}


The down quark mass-squared matrix is $\tilde{\cal M}_d^2$, 
\begin{equation}
\label{a3}
{\mathcal L} \supset \left(\tilde{Q}_3^{b*},\tilde{D}^c_3,\tilde{D}^c_4,
\tilde{D}_H^{c*}, \tilde{Q}_4^{b*}, \tilde{Q}_H^t \right)
\tilde{\mathcal M}_d^2
\left(\begin{array}{c}
\tilde{Q}_3^b \\ \tilde{D}^{c*}_3\\ \tilde{D}^{c*}_4 \\ \tilde{D}_H^c\\
\tilde{Q}_4^b \\ \tilde{Q}_H^{t*}
\end{array}\right)\,,
\end{equation}
where $\tilde{\cal M}_d^2$ is
\begin{equation}
\label{a4}
\begin{array}{cl}
(\tilde{\mathcal M}_d^2)_{11}& = 
M_Q^2-\frac{m_Z^2+2m_W^2}{6}\cos 2\beta+m_b^2+|m^d_{34}|^2\,, \\ 
(\tilde{\mathcal M}_d^2)_{12}& = (m_0-\mu\tan\beta)m_b^*\,, \\ 
(\tilde{\mathcal M}_d^2)_{13}& = (m_0-\mu\tan\beta)m^{d*}_{34}\,, \\ 
(\tilde{\mathcal M}_d^2)_{14}& = \mu^D m^{d*}_{34}\,, \\
(\tilde{\mathcal M}_d^2)_{15}& = m_b^*m^d_{43}+m^{d*}_{34}m^d_{44}\,, \\ 
(\tilde{\mathcal M}_d^2)_{16}& =  0\,, \\
(\tilde{\mathcal M}_d^2)_{21}& =(m_0-\mu\tan\beta)m_b\,, \\ 
(\tilde{\mathcal M}_d^2)_{22}& =
M_D^2-\frac{m_Z^2-m_W^2}{3}\cos 2\beta+m_b^2+|m^d_{43}|^2\,, \\
(\tilde{\mathcal M}_d^2)_{23}& = m_bm^{d*}_{34}+m^d_{43}m^{d*}_{44}\,, \\
(\tilde{\mathcal M}_d^2)_{24}& = 0 \,, \\
(\tilde{\mathcal M}_d^2)_{25}& = (m_0-\mu\tan\beta)m^d_{43}\,, \\ 
(\tilde{\mathcal M}_d^2)_{26}& = \mu^{Q*}m^d_{43}\,, \\ 
(\tilde{\mathcal M}_d^2)_{31}& =(m_0-\mu\tan\beta)m^d_{34}\,, \\ 
(\tilde{\mathcal M}_d^2)_{32}& = m_b^*m^d_{34}+m^{d*}_{43}m^d_{44}\,, \\
(\tilde{\mathcal M}_d^2)_{33}& =
|\mu^D|^2+M_D^2-\frac{m_Z^2-m_W^2}{3}\cos2\beta+|m^d_{34}|^2+|m^d_{44}|^2\,,\\
(\tilde{\mathcal M}_d^2)_{34}& = B^D\mu^D\,, \\
(\tilde{\mathcal M}_d^2)_{35}& = (m_0-\mu\tan\beta)m^d_{44}\,, \\ 
(\tilde{\mathcal M}_d^2)_{36}& = \mu^{Q*}m^d_{44}\,, \\
(\tilde{\mathcal M}_d^2)_{41}& = \mu^{D*}m^d_{34}\,, \\
(\tilde{\mathcal M}_d^2)_{42}& = 0\,, \\
(\tilde{\mathcal M}_d^2)_{43}& = B^{D*}\mu^{D*}\,, \\ 
(\tilde{\mathcal M}_d^2)_{44}& =
|\mu^D|^2+M_{DH}^2+\frac{m_Z^2-m_W^2}{3}\cos 2\beta\,, \\
(\tilde{\mathcal M}_d^2)_{45}& = \mu^{D*}m^d_{44}\,, \\
(\tilde{\mathcal M}_d^2)_{46}& = 0\,, \\
(\tilde{\mathcal M}_d^2)_{51}& = m_bm^{d*}_{43}+m^d_{34}m^{d*}_{44}\,, \\ 
(\tilde{\mathcal M}_d^2)_{52}& = (m_0-\mu\tan\beta)m^{d*}_{43}\,, \\
(\tilde{\mathcal M}_d^2)_{53}& = (m_0-\mu\tan\beta)m^{d*}_{44}\,, \\
(\tilde{\mathcal M}_d^2)_{54}& = \mu^Dm^{d*}_{44}\,, \\
(\tilde{\mathcal M}_d^2)_{55}& = 
|\mu^Q|^2+M_Q^2-\frac{m_Z^2+2m_W^2}{6}\cos2\beta+|m^d_{44}|^2+|m^d_{43}|^2
\,, \\
(\tilde{\mathcal M}_d^2)_{56}& = -B^{Q*}\mu^{Q*}\,,
\end{array}
\end{equation}
\begin{equation}
\begin{array}{cl}
(\tilde{\mathcal M}_d^2)_{61}& = 0\,, \\ 
(\tilde{\mathcal M}_d^2)_{62}& = \mu^Qm^{d*}_{43}\,, \\
(\tilde{\mathcal M}_d^2)_{63}& = \mu^Qm^{d*}_{44}\,, \\
(\tilde{\mathcal M}_d^2)_{64}& = 0\,, \\
(\tilde{\mathcal M}_d^2)_{65}& = -B^Q\mu^Q\,, \\
(\tilde{\mathcal M}_d^2)_{66}& = 
|\mu^Q|^2+M_{QH}^2+\frac{m_Z^2+2m_W^2}{6}\cos 2\beta\,.  
\end{array}
\nonumber
\end{equation}


The up quark mass-squared matrix is $\tilde{\cal M}_u^2$,
\begin{equation}
\label{a5}
{\mathcal L} \supset \left(\tilde{Q}_3^{t*},\tilde{U}^c_3,\tilde{U}^c_4, 
\tilde{U}_H^{c*}, \tilde{Q}_4^{t*}, \tilde{Q}_H^b \right) 
\tilde{\mathcal M}_u^2 
\left(\begin{array}{c}
\tilde{Q}_3^t \\ \tilde{U}^{c*}_3 \\ \tilde{U}^{c*}_4\\ \tilde{U}_H^c\\ 
\tilde{Q}_4^t \\ \tilde{Q}_H^{b*} 
\end{array}\right)\,, 
\end{equation} 
where $\tilde{\cal M}_u^2$ is 
\begin{equation}
\label{a6}
\begin{array}{cl}
(\tilde{\mathcal M}_u^2)_{11}& =
M_Q^2+\frac{4m_W^2-m_Z^2}{6}\cos 2\beta+m_t^2+|m^u_{34}|^2\,, \\
(\tilde{\mathcal M}_u^2)_{12}& = (m_0-\mu\cot\beta)m_t^*\,, \\ 
(\tilde{\mathcal M}_u^2)_{13}& = (m_0-\mu\cot\beta)m^{u*}_{34}\,, \\
(\tilde{\mathcal M}_u^2)_{14}& = \mu^U m^{u*}_{34} \,, \\
(\tilde{\mathcal M}_u^2)_{15}& = m_t^*m^u_{43}+m^{u*}_{34}m^u_{44}\,, \\
(\tilde{\mathcal M}_u^2)_{16}& = 0\,, \\
(\tilde{\mathcal M}_u^2)_{21}& = (m_0-\mu\cot\beta)m_t\,, \\
(\tilde{\mathcal M}_u^2)_{22}& = 
M_U^2 + \frac{2}{3}(m_Z^2-m_W^2)\cos 2\beta + m_t^2 + |m^u_{43}|^2\,, \\
(\tilde{\mathcal M}_u^2)_{23}& = m_tm^{u*}_{34}+m^u_{43}m^{u*}_{44}\,,\\
(\tilde{\mathcal M}_u^2)_{24}& = 0\,, \\
(\tilde{\mathcal M}_u^2)_{25}& = (m_0-\mu\cot\beta)m^u_{43}\,, \\
(\tilde{\mathcal M}_u^2)_{26}& = \mu^{Q*}m^u_{43}\,, \\
(\tilde{\mathcal M}_u^2)_{31}& = (m_0-\mu\cot\beta)m^u_{34}\,, \\
(\tilde{\mathcal M}_u^2)_{32}& = m_t^*m^u_{34}+m^{u*}_{43}m^u_{44}\,, \\
(\tilde{\mathcal M}_u^2)_{33}& = 
|\mu^U|^2+M_U^2+\frac{2}{3}(m_Z^2-m_W^2)\cos2\beta+|m^u_{34}|^2+|m^u_{44}|^2
\,, \\
(\tilde{\mathcal M}_u^2)_{34}& = B^U\mu^U\,, \\ 
(\tilde{\mathcal M}_u^2)_{35}& = (m_0-\mu\cot\beta)m^u_{44}\,, \\
(\tilde{\mathcal M}_u^2)_{36}& = \mu^{Q*}m^u_{44}+\mu^U m_H^{u*}\,, \\
(\tilde{\mathcal M}_u^2)_{41}& = \mu^{U*}m^u_{34}\,, \\
(\tilde{\mathcal M}_u^2)_{42}& = 0\,, \\
(\tilde{\mathcal M}_u^2)_{43}& = B^{U*}\mu^{U*}\,, \\
(\tilde{\mathcal M}_u^2)_{44}& = 
|\mu^U|^2 + M_{UH}^2 - \frac{2}{3}(m_Z^2-m_W^2)\cos2\beta+|m_H^u|^2\,,\\
(\tilde{\mathcal M}_u^2)_{45}& = \mu^{U*}m^u_{44} + \mu^Qm_H^{u*}\,, \\
(\tilde{\mathcal M}_u^2)_{46}& = (m_0-\mu\tan\beta)m_H^{u*}\,, \\
(\tilde{\mathcal M}_u^2)_{51}& = m_tm^{u*}_{43}+m^u_{34}m^{u*}_{44}\,,\\
(\tilde{\mathcal M}_u^2)_{52}& = (m_0-\mu\cot\beta)m^{u*}_{43}\,, \\
(\tilde{\mathcal M}_u^2)_{53}& = (m_0-\mu\cot\beta)m^{u*}_{44}\,, \\
(\tilde{\mathcal M}_u^2)_{54}& = \mu^Um^{u*}_{44}+\mu^{Q*}m_H^u\,, \\
(\tilde{\mathcal M}_u^2)_{55}& = 
|\mu^Q|^2+M_Q^2+\frac{4m_W^2-m_Z^2}{6}\cos2\beta+|m^u_{44}|^2+|m^u_{43}|^2
\,, \\
(\tilde{\mathcal M}_u^2)_{56}& = B^{Q*}\mu^{Q*}\,, 
\end{array}
\end{equation}
\begin{equation}
\begin{array}{cl}
(\tilde{\mathcal M}_u^2)_{61}& = 0\,, \\
(\tilde{\mathcal M}_u^2)_{62}& = \mu^Qm^{u*}_{43}\,, \\
(\tilde{\mathcal M}_u^2)_{63}& = \mu^Qm^{u*}_{44}+\mu^{U*}m_H^u\,, \\
(\tilde{\mathcal M}_u^2)_{64}& = (m_0-\mu\tan\beta)m_H^u\,, \\
(\tilde{\mathcal M}_u^2)_{65}& = B^Q\mu^Q\,, \\
(\tilde{\mathcal M}_u^2)_{66}& = 
|\mu^Q|^2+M_{QH}^2+|m_H^u|^2-\frac{4m_W^2-m_Z^2}{6}\cos 2\beta\,.  
\end{array}
\nonumber
\end{equation}


As we have expected, taking the $3-4$ mixing mass to be small 
$m_{34}^{l(u,d)}\to 0$, the third generation also decouples from the two 
extra generations.

\end{document}